# Probability and expected frequency of breakthroughs – a robust method of research assessment based on the double rank property of citation distributions.

Alonso Rodríguez Navarro[a,b]*, Ricardo Brito[b]

[a] *Departamento de Biotecnología-Biología Vegetal, Universidad Politécnica de Madrid, Avenida Puerta de Hierro 2, 28040, Madrid, Spain*
[b] *Departamento de Estructura de la Materia, Física Térmica y Electrónica and GISC, Universidad Complutense de Madrid, Plaza de las Ciencias 3, 28040, Madrid, Spain*

\* Corresponding author  e-mail address: alonso.rodriguez@upm.es

In research policy, effective measures that lead to improvements in the generation of knowledge must be based on reliable methods of research assessment, but for many countries and institutions this is not the case. Publication and citation analyses can be used to estimate the part played by countries and institutions in the global progress of knowledge, but a concrete method of estimation is far from evident. The challenge arises because publications that report real progress of knowledge form an extremely low proportion of all publications; in most countries and institutions such contributions appear less than once per year. One way to overcome this difficulty is to calculate probabilities instead of counting the rare events on which scientific progress is based. This study reviews and summarizes several recent publications, and adds new results that demonstrate that the citation distribution of normal publications allows the probability of the infrequent events that support the progress of knowledge to be calculated.

# 1. Introduction

Countries, public institutions, and private companies require reliable methods of research assessment to determine the profitability of their research investments and to implement research policies that improve their research systems. Despite this obvious requirement, the research policy of many countries is implemented without a rigorous assessment of the profitability of their research investments or, which is even worse, on the basis of misleading judgments and preconceptions (Rodriguez-Navarro and Narin 2018). More interestingly, most analyses of the role of R&D in the knowledge-based economy are simply based on the share of global R&D investments or the number of scientific publications (e.g., UNESCO 2016). To be correct, these analyses would require that in all countries the success of research was proportional to research investments and that all publications were similarly valuable. Neither hypothesis has been demonstrated. Furthermore, these assumptions contradict both that breakthrough papers support knowledge progress (Merton 1965) and that some papers receive many more citations than others. Consequently, it seems that the accuracy of econometric studies about the benefits of R&D investment would improve with better research assessments.

Consistent with the crucial importance of measuring research, a large number of research indicators have been proposed (Waltman 2016). In the last 20-30 years these research indicators have grown in number like a Cambrian explosion (van-Noorden 2010) or a metric tide (Wilsdon et al. 2015); this metric obsession has been severely criticized (Lawrence 2007). In fact, although many of these metrics provide reliable rankings for universities and research institutions, the indicators used to perform the rankings lessen the differences between actors, up to the point that the best research universities seem to be less cost-efficient than average universities (Rodríguez-Navarro 2012). Obviously, the use of these indicators is likely to lead to misleading judgments.

The most solid research performance indicators are based on the share of papers in top percentiles of the citation distribution of all publications in a research field (Bornmann 2013). Top percentile indicators are included by the National Science Board in its reports on Science & Engineering indicators (National Science Board 2018, 2016); they are also used in some research institution rankings (e.g., the Leiden Ranking and SCIMAGO); and the European Commission is starting to use them (European Commission 2017). However, although robust from a statistical point of view (Bornmann 2013), percentile-based assessments require the selection of a certain percentile. In most cases, this selection is done arbitrarily (Schreiber 2013) and percentiles selected are too high considering the low proportion of landmark publications (Bornmann et al. 2018).

The causes of the puzzling state of research assessments arise from the intangibility of knowledge, which is the product of research that society needs and the one that should be measured. Therefore, the best approach to performing research assessments is based on the analysis of scientific publications, taking these as a proxy of knowledge production. Publications are a clear output of all research systems and are tangible, countable, and their importance can be analyzed in many ways. The issue here is that publications that report discoveries or breakthroughs form an extremely low proportion of all publications, which implies that knowledge advancement cannot be assessed simply by counting the number of publications. As a matter of fact, the research process that ends in the achievement of a scientific advance is full of intermediate publications that are a necessary byproduct of the research process. These publications are relevant for researchers in their quest for breakthroughs and scientific advances but they do not report real scientific advances (Rodríguez-Navarro 2012). Therefore, as publications reporting important achievements are very infrequent (Bornmann et al. 2018), to count the publications that report important achievements is an impossible task. On the other hand, the ratio between the number of important achievements and the number of "normal" research publications, using a Kuhnian terminology (Kuhn 1970), is very high and highly variable across countries and institutions. Consequently, the number of achievements cannot be calculated as a simple proportion of the total number of publications.

A solution to this challenge is to calculate the probability of countries and institutions achieving one of the infrequent events that promotes the advancement of world science. This is possible and this report reviews and summarizes several recent publications, and adds new data that demonstrate that the citation distribution of normal publications allows this probability to be calculated. We also describe other related indicators and their methods of calculation, simplifying and providing graphical support to the mathematical descriptions in order to make this review accessible to a wide audience.

## 2. The probability of infrequent breakthroughs – an indicator for research assessment

As just described, the conundrum for research assessment is to calculate the contribution of a country or institution to the rare events that mediate the advancement of science, because these rare events cannot be counted at country or institution level, with very few exceptions. An alternative is to calculate the probability that a publication from a country or institution will report one of the important achievements or breakthroughs that occur annually in a particular field of research. It does not matter whether this probability is very small, provided that it can be calculated based on the total number of publications.

For this purpose, a basic step is to assign scientific importance to each publication. This assignment has been extensively studied in scientometrics and the normal procedure is to use the number of citations, because this number is highly correlated with scientific relevance (De Bellis 2009).

Some criticisms of the use of citation-based research assessments (MacRoberts and MacRoberts 2018) do not sufficiently take into account the conceptual and statistical evidence that supports the method (reviewed by (De-Bellis 2009)). More recently, and consistently with the previous literature, the institutions that are associated with Nobel Prize laureates have been identified (Schlagberger et al. 2016), and it is simple to ascertain that these institutions have the highest positions in citation-based rankings (Leiden Ranking or SCIMAGO). Similarly, the number of Nobel achievements correlates with citation indicators across countries and institutions (Rodríguez-Navarro 2011, 2016; Rodriguez-Navarro and Narin 2018). It is obvious that the procedure of awarding Nobel Prizes is based on scientific relevance and not on the number of citations, which implies that scientific relevance and number of citations are correlated.

In summary, the correlation between the number of citations and scientific relevance has strong support, and this has been built over many years by many authors, using different methods. However, it is worth highlighting that correlation implies that deviations at datum point level might be high, as this is intrinsic to the statistical concept of correlation. Therefore, from a formal point of view, research assessments based on citations cannot be used for individual papers or researchers, and nor can they be used at low aggregation levels. Just as an example it can be mentioned that some Nobel Prize discoveries had difficulties for publication and received little attention and citations for some time (Campanario 2009). An example of the opposite exists in bibliometrics with the $h$ index. This index (Hirsch 2005) is an interesting curiosity but has little relevance for evaluation purposes (Rodríguez-Navarro 2012; Bornmann and Leydesdorff 2018; Teixeira da Silva and Dobránski 2018), a fact that is in contradiction with its very high level of citations.

An additional consideration is that, to calculate the aforementioned probability by citation analysis, the existence of the correlation between citations and scientific relevance is a necessary but not a sufficient condition. Another condition is that papers that report important breakthroughs are not serendipity events that appear without any relationship to "normal" research publications. Fortunately for research assessment, this second condition is probably fulfilled because highly cited papers belong to a heavy-tailed citation distribution that can be studied by statistical methods (Glänzel 2013).

To conclude, the calculation of the probability that a paper published in a country or institution is one of the infrequent breakthrough-reporting papers published each year in a research field seems a possible objective; the challenge is to find the method of calculation.

## 3. Lognormal citation distributions

The first question to answer before trying to calculate the probability of publishing a breakthrough paper by citation analysis is how citations are distributed; the answer is that a large number of studies have shown that citation distributions follow lognormal functions (Redner 2005; Radicchi et al. 2008; Stringer et al. 2010; Evans et al. 2012; Thelwall and Wilson 2014a, 2014b; Shen et al. 2018). In addition to these mathematical studies, a further support for the lognormal distribution of citations comes indirectly from studies of the skewedness of these distributions.

The best-established of the methods used to quantify this skewedness is the *characteristic scores and scales, CSS* (Schubert et al. 1987; Glänzel and Schubert 1988), which distributes publications in several sets according to their number of citations: (i) below the mean number of citations, $m_1$; (ii) between $m_1$ and $m_2$, where $m_2$ is the mean number of citations for articles with their number of citations above $m_1$; (iii) between $m_2$ and $m_3$, where $m_3$ is the new mean of the articles with their number of citations above $m_2$. If the process is repeated three or four times, many CSS studies find that the distributions of papers in these sets are, approximately, 70-21-9% and 70-21-6-3% (Glänzel 2007; Albarrán et al. 2011; Albarrán and Ruiz-Castillo 2011; Li et al. 2013; Albarrán et al. 2015; Perianes-Rodriguez and Ruiz-Castillo 2015; Ruiz-Castillo and Waltman 2015; Bornmann et al. 2017; Viiu 2017). Recently, two independent studies have shown that this CSS distribution occurs because the citation distributions are lognormal (Rodríguez-Navarro and Brito 2018a; Viiu 2018).

Lognormal distributions, which are defined by two parameters, $\mu$ and $\sigma$, are highly right-skewed and the mean is in most cases is much higher than the mode (Aitchison and Brown 1963). In many cases they are referred to as having a heavy right tail. Figure 1 illustrates this characteristic graphically showing a lognormal distribution, $\mu = 1.7$ and $\sigma = 1.0$, that simulates an actual case of citation distribution with 150,000 papers (Rodríguez-Navarro and Brito 2018a). Figure 1a shows the distribution of the simulated publications having up to 20 citations, 135,997 publications, which implies that there are 14,003 publications (9.3%) with 21 or more citations that are not recorded in the figure. By increasing the number of citations up to 50 (Fig. 1b; 148,028 publications) the number of publications with 51 or more citations, which are not recorded, decreases to 1,972 publications (1.3%), a number that is high considering that the mean number of citations of all (simulated) publications is 9.03 and the mode is 2 (Fig. 1a). This simulation shows that the publications considered in an indicator that is frequently used for research assessment, the top 1% of most cited papers (1,500 papers with 56 or more citations), belong entirely to the long right tail that is not recorded in Fig. 1b. In actual lognormal citation distributions the values of $\mu$ and $\sigma$ are highly correlated and, on average, $\mu$ is 1.8 times higher than $\sigma$ (Rodríguez-Navarro and Brito 2018a). Higher values for $\mu$, and consequently of $\sigma$, imply that the country or institution has a higher probability of publishing highly cited papers.

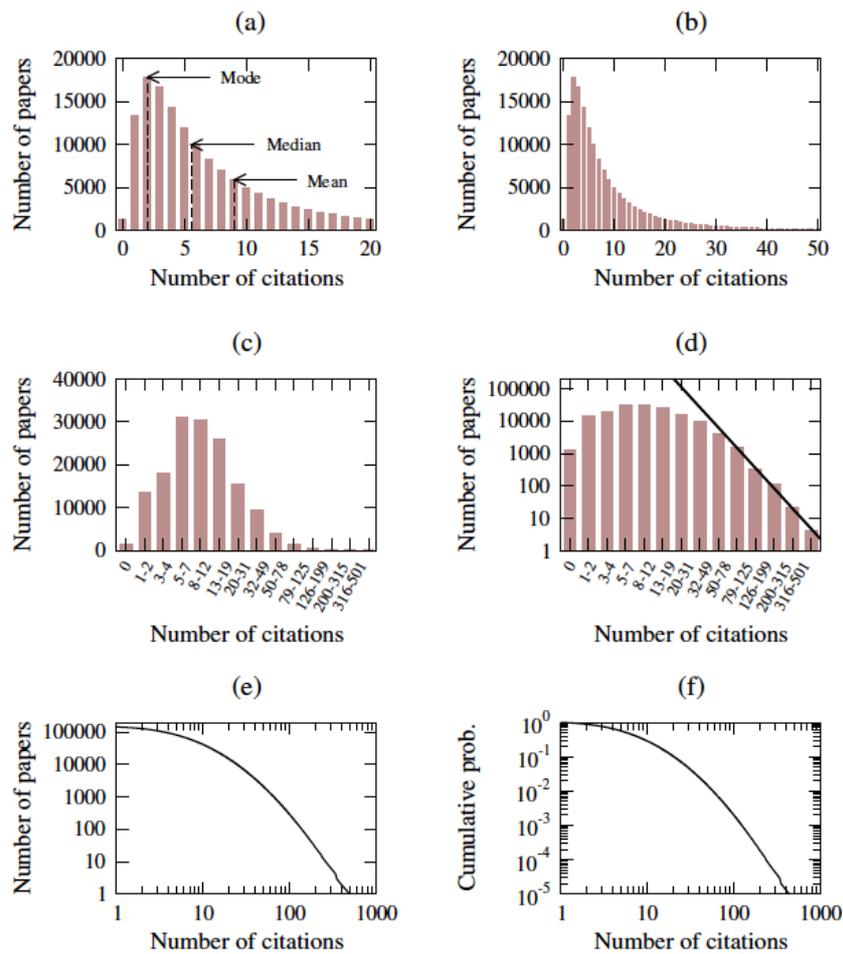

*Fig. 1. Different ways of plotting lognormal citation distributions highlight their most relevant characteristics. (a) and (b), histograms up to 20 and 50 citations; (c) and (d), logarithmic binning with normal or log-transformed frequency axis; (e) and (f), rank-frequency and cumulative probability plots, both in double logarithmic scales. Synthetic data, $\mu = 1.7$, $\sigma = 1.1$, N = 150,000 (Rodríguez-Navarro and Brito 2018a).*

To show in a single figure all the data of these skewed distributions, including the data omitted in Fig. 1b, the use of logarithmic binning is a solution. Using this binning approach, Fig. 1c,d show the same data as those presented in Fig. 1a,b, but they include all the data points. In Fig. 1c, the skewed distribution looks like a normal distribution, which necessarily occurs in a lognormal distribution. In this figure the left half of the distribution is not exactly symmetric with the right half. This occurs because of the discrete nature of citations and the consequent impossibility of making perfect logarithmic bins for low numbers of citations (1-2, and 3-4). Using a logarithmic scale for the number of papers as well, the shape of the distribution changes and the right tail approximates to a straight line (Fig. 1d).

Another way of representing the citation lognormal distribution is to use a rank-frequency plot (Newman 2005). To construct this plot, all papers are ranked according to their number of citations, starting with the paper that has the highest number of citations and numbering them starting from 1. The rank number is then plotted as a function of the number of citations as shown in Fig. 1e. In this type of plot, dividing the rank number by the total number of papers gives the *cumulative probability function*, which indicates the probability that a paper has received at least a given number of citations (Fig. 1f).

This *cumulative probability function* could be the basis of the evaluation procedure required in the previous section. However, the implementation of the method is not straightforward. As discussed in Section 5 below, the use of the *cumulative probability function* of the lognormal distribution presents several difficulties that need to be overcome.

**4. The pseudo-power-law tail**

An interesting characteristic of the right tails of lognormal citation distributions is that if log-log scales are used, in both logarithmic binning (Fig. 1d) and rank-frequency plots (Fig. 1e), the data points can be fitted to a straight line (Zip's law; a review about this tail is in (Katz 2016)). This graphical property is characteristic of power laws defined by the following equation:

$$y = a\, x^{-\alpha} \qquad [1]$$

However, the power law that best fits the tail of a lognormal citation distribution (Section 2) is coincident with the lognormal data points in a large section but clearly deviates in the last points on the right, as shown in Fig. 2. By adding an exponential cut-off to this power law, the deviations from the lognormal data points are decreased (Katz 2016) but a perfect fit is not possible because the tail does not obey a power law.

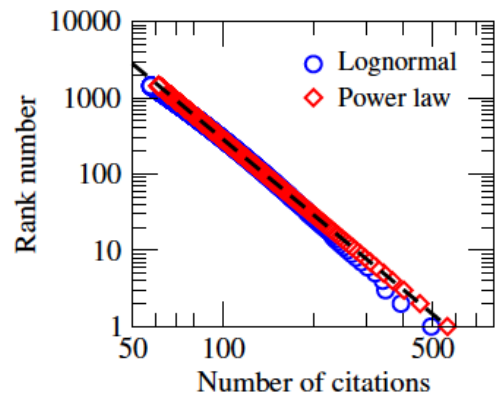

*Fig. 2. The right tail of lognormal citation distributions approximate to power laws. The figure shows the 1,500 data points in the right tail of the lognormal distribution displayed in Fig. 1. The power law data points that are shown were obtained after fitting a power law function to the lognormal data points from 400 down to 50. The last 20 points clearly deviate from the straight line of the real power law in the log-log plot.*

Accepting the deviations, the power law function that best fits the tail of the lognormal citation distribution can be used to calculate the cumulative frequency with which a country or institution publishes very highly cited papers. To perform a research assessment in this way, the only requirement is that an appropriate number of citations is selected. This approach has been validated in terms of Nobel Prizes garnered by the USA and the EU (Rodríguez-Navarro 2016). However, independently of accuracy problems, the method cannot be used for small countries or institutions because the number of data points of the upper tail that can be fitted to a power law is too low. This number is approximately 1% or less of all publications (Katz 2016 and references therein), and with such a low number of data points a reliable fitting of the pseudo power law is impossible and the method is non-viable.

A conceptual issue associated with a research assessment based only on the top 1% of cited publications is the relationship between these top-class publications and all other publications. If the two groups were independent, the bulk of publications (99%) might be unnecessary for the progress of science. This issue connects with the question of whether only an elite among scientists is boosting the progress of science, with the contribution of other researchers being negligible (Cole and Cole 1972; Bornmann et al. 2010).

Fortunately, this discussion can be settled because, as we have mentioned in Section 3, citation distributions, including both the 1% right tail and the remaining 99% of the data points, obey a lognormal function. This implies that in research all papers count, although some count more than others. Certainly, science progresses on the shoulders of giants (Merton 1965), but the lognormal citation distribution suggests that the giants are not isolated from the mass-man masses of scientists (Ortega y Gasset 1999). In contrast with the old dichotomous scheme of giants and the mass of scientists, it seems that research is performed by a continuity of scientists from giant to the least important and that all are needed, although the proportions vary across countries and institutions.

**5. The percentile-based double rank analysis**

As already mentioned, to calculate the probability of an institution or country publishing very highly cited papers, the use of the *cumulative probability function* is a good option that can be achieved after fitting a lognormal function to the paper citation data (Section 3). However, despite apparently favorable prospects, research assessments that rely on the lognormal distribution of citations present a substantial difficulty because an isolated citation distribution lacks a reference point to fix the position of success.

One way to avoid this difficulty is to compare the lognormal distribution of the papers published by a country or institution to the lognormal distribution of the world's papers in the research field. A simple method of comparison takes it for granted that the global distribution is made up of the combination of multiple local lognormal distributions, which implies that all global publications also belong to an institution or country. Thus, a convenient way of comparing local and global publications can therefore use these two rankings (local and global). With this purpose, as described above, we rank local and global papers from the highest to the lowest number of citations. Then, all the local papers have two rank position numbers because they are in both the local and the global lists. The local rank position number can then be expressed as a function of the global rank position number; this function should reveal the differences between countries and institutions, and be sufficient for calculating the probability of publishing highly cited papers.

This double ranking approach was investigated with a series of synthetic data that simulated local and global citation distributions (Rodríguez-Navarro and Brito 2018a). Figure 3 shows three sets of data from that study that simulate all world publications in chemistry, 151,000 papers and $\mu$ = 1.7, and two research institutions, one, s1, producing excellent research with many highly cited papers, which implies a high $\mu$ ($\mu$ =

2.4), and the other, s7, much less competitive, showing a low number of highly cited papers and a lower $\mu$ ($\mu$ = 1.5). Figure 3a shows the binning distribution of the two institutions, with the right tails truncated at 25 citations; Fig. 3b shows the rank-frequency plots of all the data points of the same distributions. Obviously, when the papers from the simulated excellent (s1) and poor (s7) research institutions are located in the global set, they are differently distributed. At high citation levels, the s1 data points (Fig. 3c) are more frequent than the s7 data points (Fig. 3d). Thus, the most highly cited paper from the s1 set (local rank number 1) ranks number 4 in the global set, and the most highly cited paper from the s7 set (local rank number 1) ranks number 730 in the global set.

For research assessment, we need to know the function that links the local and the global rank numbers and that describes the distribution of the local papers among the global papers. This function was studied and found to be a power law (Rodríguez-Navarro and Brito 2018a).

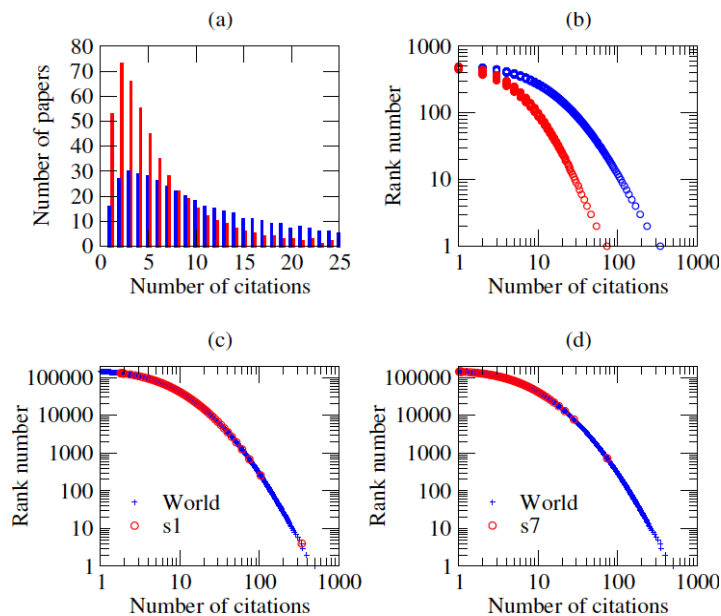

*Fig. 3. Plots of synthetic data that follow two lognormal citation distributions, s1 and s7, which are part of a third lognormal distribution that simulates the world citation distribution. (a), histograms of the two lognormal distributions up to 25 citations; (b), rank-frequency plots of the two distributions; (c) and (d), positions of the s1 and s7 data points in the world distribution. Synthetic data: s1, $\mu$ = 2.4, $\sigma$ = 1.1, N = 500; s7, $\mu$ = 1.5, $\sigma$ = 0.9, N = 500; world, $\mu$ = 1.7, $\sigma$ = 1.1, N = 151,000 (Rodríguez-Navarro and Brito 2018a). To show the distribution more clearly, in (c) and (d) only the first data point in each set of 10 is highlighted.*

To further study this power law for research assessments, the world rank numbers in the double rank plot can be expressed in terms of world percentiles, and the local rank numbers as the number of papers in each percentile. The rationale for this variant of the double rank method is the convenience of using a research assessment method that incorporates all the advantages of the percentile apportionment (Bornmann 2013). This transformation does not affect the double rank function because using the top percentiles of global papers is another way of expressing the global rank number and the number of local papers in the top percentiles corresponds to the local rank numbers of these papers (Brito and Rodríguez-Navarro 2018a). In consequence, the distribution of the papers in a research field in the top percentiles follows a power law. Figure 4 shows this fact with both synthetic data and empirical data taken from (Brito and Rodríguez-Navarro 2018a). This power law function can be written in the following way

$$N(x) = A\, x^\alpha \qquad [2]$$

where *x* is the top percentile and *N(x)* is the number of papers in this percentile. This equation can also be written as a function of the total number of papers of the country, which is the *cumulative frequency function*

$$N(x) = N\left(\frac{x}{100}\right)^\alpha \qquad [3]$$

where *N* is the total number of papers. Then, the *cumulative probability function* is

$$P(x) = \left(\frac{x}{100}\right)^\alpha \qquad [4]$$

Simple mathematic operations allow the parameters of Eq. [1] to be expressed in terms of the top 1% and top 10% of most cited papers (*N*(1) and *N*(10), which are denoted by $P_{top\,1\%}$ and $P_{top\,10\%}$ in the widely used Leiden Ranking notation)

$$A = P_{top\,1\%} \qquad [5]$$
$$\alpha = \lg(P_{top\,10\%}/P_{top\,1\%}) \qquad [6]$$

Systematic analyses of percentile-based double rank distributions calculated from synthetic lognormal distributions having parameters *μ* and *σ* within the range of values that occur in actual citation distributions demonstrate that the power law function mentioned above always appears. Slight deviations from the power law occur only for local lognormal distributions with *μ* values much higher than the global distribution and only at high percentiles (Brito and Rodríguez-Navarro 2018a).

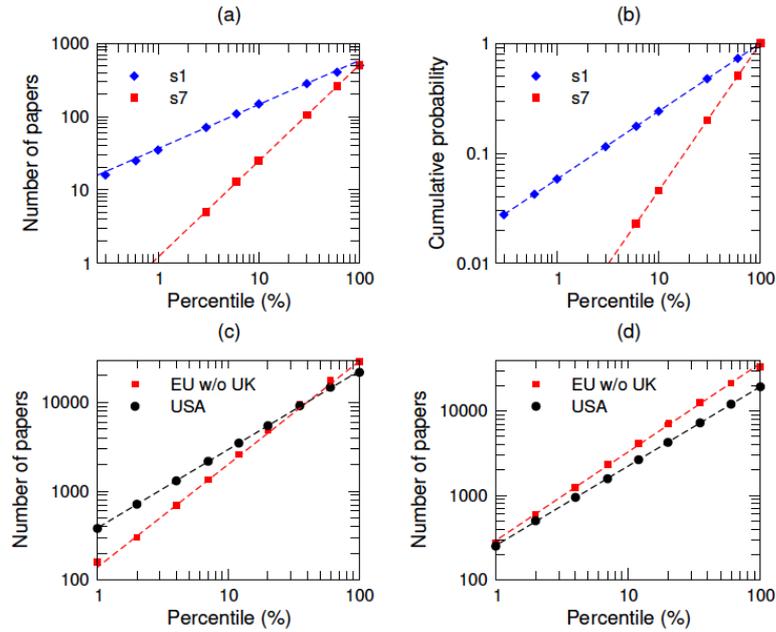

*Fig. 4. Paper distributions and cumulative probabilities in world percentiles obey power law functions. (a) and (b): synthetic data, the two lognormal distributed sets of data, s1 and s7, and the simulated world lognormal set of data are the same as those used in Fig. 3. (c) and (d) empirical data for the USA and the EU, excluding UK, in fast-evolving (c) and slow-evolving (d) technological fields (reproduction of Fig. 1 in Rodríguez-Navarro and Brito 2018b).*

Exceptionally the data points at low percentiles deviate from the main trend of the distribution; an example is publications on graphene in the EU excluding the UK (our unpublished results). It seems that the aggregation of a low number of highly and lowly competitive countries causes this deviation. The mathematical explanation is that the assembling of a low number of lognormal distributions with very different $\mu$ values does not generate a lognormal distribution.

It is worth noting that, in actual assessments, the number of citations that is recorded for each paper depends on the research field (e.g., cancer papers receive many more citations that mathematical papers) and on the citation window (e.g., using a citation window of five years, the number of citations will be approximately five times higher than using a citation window of one year). These circumstances affect to the world, country, and institution distributions identically, which implies that the double rank analysis is independent of them and provides comparable results in all cases.

The accuracy with which the double rank power law function fits the synthetic data strongly suggests that there should be an analytical demonstration of this power law. However, no demonstration has been found so far, and the focus is currently on the statistics that support the results of the empirical data.

**6. Fitting of the percentile-based double rank power law is statistically robust**

Consistent with the results obtained with synthetic data, the real citation-based distributions of country and institution publications in the percentiles of global publications follow a power law that shows negligible deviations (Fig. 4c and d). Only at high percentiles, mainly 100%, can the deviations be appreciable. For evaluation purposes the accuracy of the function fitting is important because, as described in Section 2, the double rank approach is directed to the specific purpose of obtaining extrapolated values of the variable outside of the range of empirical values.

*6.1. A case study: USA and EU research in fast evolving technologies*

To study the accuracy with which a power law function describes the percentile-based double rank distributions of publications, we will use data that have already been published for the USA and the EU in fast-evolving technologies (Rodríguez-Navarro and Brito 2018b). In both the USA and the EU, the data points at the 100 percentile deviate slightly from the power law, although these deviations are imperceptible in log-log plots (Fig. 4c). To perform a more rigorous quantitative analysis of the data, two steps have to be considered: the fitting approach and the assessment of the goodness-of-fit. Concerning the fitting method of the double rank power law, two atypical characteristics have to be considered. In the first place, the exponent of the power law is positive, when in most cases studied the power law has a negative exponent, as in Eq. [1] (e.g., Goldstein et al. 2004; Bauke 2007; Clauset et al. 2009). A negative exponent implies that when the intensity of the phenomenon increases, the probability of it happening decreases. In our case, the response is just the opposite: when the top percentile increases the probability also increases (Fig. 4). In the second place, the fitting procedure must be specifically designed for the purpose that is sought, which is extrapolation from empirical data. In most cases, maximum likelihood estimation provides the best fitting parameters (Goldstein et al. 2004; Bauke 2007; Clauset et al. 2009), but in the double rank approach the best fitting must be investigated. The purpose is to find a method that provides maximum accuracy at low percentiles when maximum deviations of the data are expected at high percentiles, which implies that the best method is the one that is the least sensitive to these deviations.

To find the best method of fitting, we compared the results of the three methods that are most frequently used: maximum likelihood (ML), Levenberg Marqardt (LM) nonlinear fitting, and linear regression with log-transformed data. For this analysis, we took advantage of the aforementioned deviations at high percentiles by performing fits including and excluding the 100% and 60% data points and investigating the goodness-of-fit by the *p*-values calculated by using the $\chi^2$ Statistics (Press et al. 1989). In addition we calculated the values of the $P_{top\ 0.01\%}$ indicator, to investigate the method that was least sensitive to the deviations at high percentiles. The results, which are summarized in Table 1, lead to clear conclusions.

Table 1. Fits to power law functions of the percentile-based double rank data for the USA and the EU excluding the UK in fast evolving technologies in 2014. Fitting has been performed both including all data points and excluding the 100 percentile[a]

| | All percentiles | | | | Excluding 100% | | | | Excluding 100% and 60% | | | |
|---|---|---|---|---|---|---|---|---|---|---|---|---|
| | *A* | *α* | *p*[b] | $P_{0.01}$[c] | *A* | *α* | *p*[b] | $P_{0.01}$[c] | *A* | *α* | *p*[b] | $P_{0.01}$[c] |
| USA | | | | | | | | | | | | |
| ML[d] | 402.8 | 0.873 | <0.01 | 7.25 | 379.0 | 0.895 | 0.995 | 6.14 | 379.7 | 0.894 | 0.998 | 6.18 |
| LM | 443.6 | 0.849 | <0.01 | 8.90 | 375.9 | 0.898 | 0.994 | 6.03 | 372.7 | 0.900 | 0.976 | 5.89 |
| LR | 386.1 | 0.885 | <0.01 | 6.55 | 382.3 | 0.892 | 0.980 | 6.28 | 382.7 | 0.891 | 0.991 | 6.32 |
| EU excluding UK | | | | | | | | | | | | |
| ML[c] | 160.9 | 1.135 | <0.01 | 0.86 | 138.5 | 1.185 | 0.33 | 0.590 | 134.3 | 1.198 | 0.50 | 0.539 |
| LM | 203.8 | 1.079 | <0.01 | 1.42 | 145.0 | 1.173 | 0.13 | 0.653 | 133.2 | 1.200 | 0.45 | 0.530 |
| LR | 145.9 | 1,171 | <0.01 | 0.70 | 143.9 | 1.171 | 0.01 | 0.656 | 144.6 | 1.167 | 0.15 | 0.670 |

[a] Source of the data (Brito and Rodríguez-Navarro 2018b). Percentiles used for fitting 1, 2, 4, 7, 12, 20, 35, 60, and 100. Abbreviations of fitting methods: maximum likelihood (ML), Levenberg Marqardt (LM), linear regression with log-transformed data (LR).
[b] *p*-value is calculated by using the $\chi^2$ Statistics (Press et al. 1989).
[c] The Leiden Ranking $P_{top\ 0.01\%}$ indicator
[d] Maximum likelihood estimations were performed as described by Bauke 2007

Considering the *p*-values, all the data points cannot be fitted to a power law (*p* < 0.05). However, only by excluding the 100 percentile, a power law function explains the empirical data for the USA (*p* ≈ 0.9) and for the EU when the power law parameters were estimated by ML and LM methods (*p* > 0.1). In the case of the linear regression of log-transformed data, a good fitting required the exclusion of the 100 and 60 percentiles (*p* = 0.15).

In contrast, attending to the $P_{top\ 0.01\%}$ indicator, the linear regression shows the highest stability, as this method gives more weight to the data points at low percentiles. In the USA, when the 100 percentile was included the value of the $P_{top\ 0.01\%}$ increased by ≈ 20% and ≈ 50% for the ML and LM fittings, respectively, and in the EU the increases were higher, ≈ 60% and ≈ 120%, respectively. In the linear fitting the increases were much smaller, 7% in the worse case. For all the reasons, we will use the linear fitting to log-log transformed data.

*6.2. Statistical robustness across research areas*

To carry out further study of the statistical robustness of the percentile-based double rank analysis across research areas, we took advantage of the large amount of data reported by the National Science Board (National Science Board 2016). This report contains the percentile distributions of papers published in 2002 and 2012 in all sciences and 13 research areas: engineering, astronomy, chemistry, physics, geosciences, mathematics, computer sciences, agricultural sciences, biological sciences, medical sciences, other life sciences, psychology, and social sciences, for the USA, the EU, China, and Japan, in five top percentiles: 1, 5, 10, 25, and 50. The 100 percentile is not recorded because the results are given as percentages of the total number of publications, and this would be 100 in all cases. In total there are 112 cases to study. Using linear regression for the reasons explained in the previous section, fittings of the data points in these 112 cases we found that four were atypical: China, astronomy, 2002; China, other life sciences, 2002; Japan, psychology, 2002 and 2012, because one data point in three cases and two data points in the other case deviate widely from the pattern of the other data points. show high goodness-of-fit. Table 2 records the values of $\alpha$ and *p*-values calculated by using the $\chi^2$ Statistics (Press et al. 1989). The $\alpha$ values show a good agreement between the two years analyzed by the National Science Board, considering that the annual variability of citation distributions is notable (Rodríguez-Navarro 2016). In all cases the *p*-values values are high, being equal or above 0.999 in 77% of the fits and, except in the aforementioned four cases, they were always higher than 0.8. These high *p*-values indicate that, in contrast to other empirical data that follow a power law, where data are generally very noisy (Bauke 2007), the percentile distributions of papers are not noisy in the countries, years, and research areas that we studied. These results lead to the conclusion that a research assessment based on the percentile distribution of publications is a robust method across research areas.

Interestingly, although the goodness-of-fits based on *p*-values do not show differences between the four country cases studied in the National Science Board report (the USA, the EU, China, and Japan), a refined analysis of the data shows that in some research areas there were two patterns of small deviations from the power law: one pattern applies to the USA, and the EU and the other applies to China and Japan (as aforementioned, maximum deviations occur at the 100 percentile). In Table 3 we show the differences between the empirical data and the data calculated from the fitted power laws excluding the 100 percentile data points. In the USA and the EU the differences are negative in the 1, 50, and 100 percentiles, while in China and Japan the differences are negative in the 5, 10, and 25 percentiles. Table 3 shows the deviations in the "all sciences" compilation, where they are clearer than in individual research areas. To interpret these deviations we have to consider that the double rank assessment analyzes the positions of local publications in the global distribution according to the number of citations—that is the global distribution is an "average" of all countries. Therefore, it seems that citation practices are not identical in western and eastern countries, at least in some research fields. However, it is worth noting that the deviations at low percentiles are small and their effect on the calculation of the probabilities of expected frequencies of highly cited papers is negligible in comparison with the variability across countries and institutions.

Table 2. Fitting to power laws of the percentile data reported by the National Science Board (2016) for the publications from USA, EU, China, and Japan in several research fields in 2002 AND 2012. Exponent of the power law and *p* values of linear fitting of log-transformed data[a]

| Field | | USA | | EU | | China | | Japan | |
|---|---|---|---|---|---|---|---|---|---|
| | | 2002 | 2012 | 2002 | 2012 | 2002 | 2012 | 2002 | 2012 |
| All | α | 0.901 | 0.876 | 1.019 | 0.961 | 1.090 | 1.034 | 1.104 | 1.038 |
| | p | >0.999 | 0.999 | >0.999 | >0.999 | >0.999 | >0.999 | >0.999 | >0.999 |
| Engineering | α | 0.859 | 0.827 | 1.010 | 0.960 | 1.126 | 1.063 | 1.052 | 1.018 |
| | p | >0.999 | >0.999 | >0.999 | 0.999 | 0.998 | 0.999 | 0.992 | 0.999 |
| Astronomy | α | 0.897 | 0.902 | 0.990 | 0.964 | | 1.071 | 1.437 | 1.856 |
| | p | >0.999 | >0.999 | >0.999 | >0.999 | 0.116 | 0.903 | 0.996 | >0.999 |
| Chemistry | α | 0.874 | 0.896 | 1.033 | 1.034 | 1.029 | 0.934 | 1.055 | 1.140 |
| | p | 0.998 | >0.999 | >0.999 | >0.999 | >0.999 | >0.999 | >0.999 | >0.999 |
| Physics | α | 0.891 | 0.839 | 0.990 | 0.947 | 1.092 | 1.060 | 1.095 | 0.955 |
| | p | >0.999 | >0.999 | >0.999 | >0.999 | 0.997 | >0.999 | >0.999 | >0.999 |
| Geosciences | α | 0.927 | 0.884 | 1.021 | 0.964 | 0.802 | 0.975 | 1.121 | 0.976 |
| | p | >0.999 | 0.998 | >0.999 | >0.999 | >0.999 | >0.999 | >0.999 | >0.999 |
| Mathematics | α | 0.917 | 0.961 | 1.039 | 1.023 | 0.978 | 0.942 | 1.019 | 1.182 |
| | p | >0.999 | >0.999 | >0.999 | >0.999 | 0.996 | >0.999 | 0.932 | >0.999 |
| Computer sciences | α | 0.865 | 0.847 | 1.085 | >0.999 | 1.324 | 1.015 | 1.310 | 1.240 |
| | p | >0.999 | 0.998 | >0.999 | 0.999 | 0.990 | 0.997 | >0.999 | 0.999 |
| Agricultural sciences | α | 0.888 | 0.898 | 0.998 | 0.928 | 0.939 | 1.099 | 1.175 | 1.255 |
| | p | >0.999 | >0.999 | >0.999 | >0.999 | 0.987 | 0.996 | 0.997 | >0.999 |
| Biological sciences | α | 0.926 | 0.888 | 1.022 | 0.945 | 1.210 | 1.107 | 1.110 | 1.015 |
| | p | >0.999 | >0.999 | >0.999 | >0.999 | >0.999 | >0.999 | >0.999 | >0.999 |
| Medical sciences | α | 0.896 | 0.878 | 1.018 | 0.933 | 1.161 | 1.116 | 1.183 | 1.055 |
| | p | >0.999 | >0.999 | >0.999 | >0.999 | >0.999 | >0.999 | 0.996 | 0.997 |
| Other life sciences | α | 0.986 | 0.965 | 0.982 | 0.950 | | 1.032 | 0.994 | 1.136 |
| | p | >0.999 | >0.999 | >0.999 | >0.999 | 0.425 | 0.999 | >0.999 | 0.998 |
| Psychology | α | 0.941 | 0.918 | 1.078 | 1.019 | 1.300 | 0.972 | | |
| | p | >0.999 | >0.999 | >0.999 | >0.999 | 0.997 | 0.953 | 0.6089 | 0.7203 |
| Social sciences | α | 0.921 | 0.927 | 1.078 | 1.001 | 1.160 | 0.844 | 1.134 | 1.046 |
| | p | >0.999 | >0.999 | >0.999 | >0.999 | 0.975 | >0.999 | >0.999 | >0.999 |

[a] *p*-value is calculated by using the $\chi^2$ Statistics (Press et al. 1989)

Table 3. Deviation of empirical and power law calculated data in the USA, the EU, China, and Japan[a]

| Percentile | USA | | | EU | | |
|---|---|---|---|---|---|---|
| | Empirical | Calculated | Difference | Empirical | Calculated | Difference |
| 1 | 1.94 | 1.99 | **-0.05** | 1.29 | 1.32 | **-0.03** |
| 5 | 8.33 | 8.15 | 0.18 | 6.30 | 6.19 | 0.11 |
| 10 | 15.40 | 14.96 | 0.44 | 12.33 | 12.05 | 0.28 |
| 25 | 33.68 | 33.39 | 0.29 | 29.37 | 29.08 | 0.29 |
| 50 | 59.29 | 61.29 | **-2.00** | 55.01 | 56.62 | **-1.61** |
| 100 | 100.00 | 112.48 | **-12.48** | 100.00 | 110.23 | **-10.23** |
| | China | | | Japan | | |
| 1 | 0.81 | 0.79 | 0.02 | 0.82 | 0.79 | 0.03 |
| 5 | 4.12 | 4.18 | **-0.06** | 4.03 | 4.18 | **-0.15** |
| 10 | 8.34 | 8.57 | **-0.23** | 8.18 | 8.58 | **-0.40** |
| 25 | 21.92 | 22.10 | **-0.18** | 22.17 | 22.21 | **-0.04** |
| 50 | 46.62 | 45.25 | 1.37 | 47.84 | 45.60 | 2.24 |
| 100 | 100.00 | 92.67 | 7.33 | 100.00 | 93.63 | 6.37 |

[a] Data taken from the National Science Board (2016), field all sciences, year 2012. Power laws were fitted to the empirical data excluding de 100 percentile by linear regression. Negative differences are in bold

To conclude this analysis it is worth noting that publication and citation in the natural and the social sciences might show significant intrinsic differences. Thus, it might be possible that in the social sciences the interpretation of the double rank assessment results is more complex than in the natural sciences. Therefore, it is remarkable that the fitting of the double rank power law is similarly accurate in both cases (Table 2).

*6.3. Practical recommendations for research assessment*

In the two previous subsections we have shown that research assessment based on the double rank analysis is a statistically robust approach that, from a mathematical point of view, can be used in all research areas. However, the accuracy of the method can be improved by following some guidelines.

It is worth highlighting once more that, although as a general rule the best method for power law function fitting is maximum likelihood (Goldstein et al. 2004; Bauke 2007; Clauset et al. 2009), the percentile-based double rank approach does not follow this general rule. In this case, attending to the minor effect on the values of the $P_{\text{top } 0.01\%}$ indicator, the linear fitting with log-transformed data was more accurate than the non-linear and maximum likelihood methods.

Finally, in previous assessments (Brito and Rodríguez-Navarro 2018b; Rodríguez-Navarro and Brito 2018b) we found that in some cases low percentiles with a low number of publications (e.g., <10) are noisy. Therefore, deleting the percentile data with a low number of publications (≈ 10) is a convenient approach to improve the results.

**7. Three main indicators can be obtained from the percentile power law function**

In Section 2 we discussed how a reasonable method of research assessment could be based on the probability that a publication from a country or institution reports one of the important achievements or breakthroughs that occur each year in a particular field of research. The percentile double rank function allows the calculation of this probability, but also the calculation of two other indicators that have other properties is also interesting for research assessment: the expected frequency of breakthrough papers and the $e_p$ index.

*7. 1. The probability that a publication from a country or institution reports a breakthrough*

On the basis of distribution of local publications in global percentiles, Eq. [4] allows this probability to be calculated once the exponent $\alpha$ has been determined by fitting of the observed data to a power law (Eq. [3]). Equation [4] provides the cumulative probability that a paper lies in a certain percentile or, in other words, that a paper receives a number of citations greater than or equal to a certain threshold.

The first step for the calculation of this probability is the selection of the top percentile. In other types of evaluations based on highly cited papers, the arbitrariness of the selection of the number of citations or percentile threshold has been discussed (Schreiber 2013). This problem of arbitrariness can be eliminated, making the selection discretionary but not arbitrary, if it is based on the judgment of experts about the number of papers that annually report breakthroughs in a specific field. The percentile to be selected is obtained by dividing this estimated number of breakthroughs by the total number of world papers. In any case, the probability indicator is valid at any percentile; policy makers might find it convenient to perform less stringent assessments in developing countries.

In two previous studies (Brito and Rodríguez-Navarro 2018b; Rodríguez-Navarro and Brito 2018b) the 0.01 percentile was used. The discretionary selection of this percentile is consistent with the later finding that fewer than 0.02% of all papers are landmark publications (Bornmann et al. 2018). The 0.01 percentile might be of general use in fields with 100,000 or more annual publications.

*7. 2. The expected frequency of publications reporting breakthroughs*

The probability described in the previous section is a size-independent indicator that provides information about the efficacy of the system but does not provide data that can be used to refer to investments or country size. For example, to determine cost-effectiveness, a size-dependent indicator is necessary. In other cases, it might be interesting to determine the contribution of certain universities to the total research performance of a country. Thus, in many cases, the use of a size-dependent indicator is necessary.

A size-dependent indicator that can be calculated from the power law is the cumulative frequency of breakthrough papers (Eq. [3]). This cumulative frequency reveals the expected number of breakthrough papers from a country or institution that contribute to the advancement of science or technology or, in other words, the share of a country or institution in the total number of global papers in the selected percentile. In the case of the 0.01 percentile the cumulative frequency can be named $P_{top\ 0.01\%}$ following the Leiden Ranking notation.

*7. 3. The $e_p$ index*

The two indicators described above vary depending on the percentile selected for the evaluation. However, the equations that define both indicators depend on a single parameter, the exponent of the power law function, Eq. [2]. This exponent reveals an intrinsic characteristic of the power law, which in turn indicates the distribution of local papers among all world papers when they are ranked by the number of citations. Because the power law exponent decreases when the competence of the research system increases, we created the $e_p$ index (**e**fficiency based on **p**ercentiles or **e**xcellence based on **p**ercentiles), which increases when the competence increases (Rodríguez-Navarro and Brito 2018b):

$$e_p = P_{top\ 1\%}/P_{top\ 10\%} \qquad [7]$$

$$e_p = 10^{-\alpha} \qquad [8]$$

where $\alpha$ is the exponent in Eq. [2]. As a reference, when the lognormal distribution of a country or institution has the same $\mu$ and $\sigma$ parameters as the world distribution, the value of the $e_p$ index is 0.1. Values of the $e_p$ index higher or lower than 0.1 imply, respectively, better or worse performances than the global average.

The $e_p$ index measures the *intrinsic efficiency* or *breakthrough potential* of a research system. Because highly cited papers cannot exist without quite a high number of infrequently cited papers, the $e_p$ index measures the intrinsic efficiency of the research system in producing papers at a given citation level from the distribution of papers at lower levels of citation (Rodríguez-Navarro and Brito 2018b). The aforementioned $P_{top\ 0.01\%}$ indicator can be easily calculate using the $e_p$ index. From Eq. (3) and (8) it is obtained that

$$P_{top\ 0.01\%} = N\ e_p^4 \qquad [9]$$

## 8. Validation of the double rank indicators

As a general rule, indicators must be validated by correlation (Harnad 2008, 2009) and an obvious problem with the Cambrian explosion of research metrics is that most of these metrics have not been validated or have only been validated against other doubtful indicators. For example the popular *h* index shows a reasonable correlation with some other indicators, such as the number of citations (van-Raan 2006), but is not an accurate index for research evaluation (Rodríguez-Navarro 2012; Bornmann and Leydesdorff 2018; Teixeira da Silva and Dobránski 2018).

Indicators calculated from the double rank function at low percentiles are conceptually identical to those calculated from the pseudo power law tail (Section 4), which have been validated in terms of the number of Nobel Prize achievements (Rodríguez-Navarro 2016). To further validate the double rank indicators, we calculated the $P_{top\ 0.01\%}$ for the USA and the EU from the data provided by the National Science Board (National Science Board 2016) in the research areas of chemistry, physics, biological sciences, and medical sciences. These results can be compared to the frequency of Nobel Prize awards in Chemistry, Physics, and Physiology or Medicine. Table 4 shows these results and the USA/EU ratios. In the period 1982–2007 these ratios show a notable annual variability (Rodríguez-Navarro 2016; Rodriguez-Navarro and Narin 2018), and the data recorded in Table 4 fall within the ranges of this variability, which implies that are in agreement with the number of Nobel Prizes garnered by the USA and the UE.

Table 4. $P_{top\ 0.01\%}$ values for the USA and the EU calculated from the data provide by the National Science Board[a]

| Field | 2002 | | | 2012 | | |
|---|---|---|---|---|---|---|
| | USA | EU | Ratio | USA | EU | Ratio |
| Chemistry | 0.0370 | 0.0085 | 4.35 | 0.0301 | 0.0086 | 3.5 |
| Physics | 0.0300 | 0.0121 | 2.48 | 0.0449 | 0.0178 | 2.52 |
| Biological Sciences | 0.0234 | 0.0088 | 2.66 | 0.0340 | 0.0190 | 1.79 |
| Medical Sciences | 0.0320 | 0.0081 | 3.95 | 0.0368 | 0.0190 | 1.94 |

[a] (National Science Board 2016)

In summary, the percentile-based double rank indicators are powerful tools for research assessment; both mathematical considerations and empirical analyses validate the results of research assessments based on this approach.

**9. Double rank indicators can substitute for peer assessments**

Because society and private investors need to know the profitability of their large investments in research, in parallel with the many proposals of methods of research assessment there has been a long-lasting discussion on the pros and cons of bibliometric-based and peer-based research assessments, as described, for example, in "The Metric Tide" (Wilsdon et al. 2015). It is worth noting that both approaches need to be validated because "even those who wish to refute metrics in favor of peer review first have to demonstrate that peer review is somehow more reliable and valid than metrics" (Harnad 2008 p. 103). The difficulties arise because "science is a part of society marked by inequality, random chance, anomalies, the right to make mistakes, unpredictability and a high significance of extreme events" (Bornmann 2017 p. 780). Consistent with these difficulties, we have already mentioned that publications recording achievements that were awarded with Nobel Prizes after several years were not originally well evaluated by peers (Campanario 2009). In summary, peer reviews are not always perfect.

Before describing how the double rank indicators fit within this discussion, it must be emphasized once more (see Section 2) that there is a great difference between evaluating individual papers or researchers and evaluating large groups of papers or researchers, as in the evaluation of countries or institutions. The basis of the difference is that no bibliometric indicator is a genuine metric of knowledge generation. Some bibliometric indicators correlate with the relevance of research outputs (Section 2), but they do not measure their research relevance; therefore, it does not matter how high the correlation is because at the data point level the deviations might be high. Therefore, the conclusion is obvious from a formal point of view: no bibliometric indicator can be used to evaluate individual publications or researchers, although, in the hands of experts, some—but not all—bibliometric indicators can help in peer-performed research assessments.

Consistent with these ideas, the double rank indicators described in the previous section have been developed only to evaluate large sets of publications from countries, institutions, or groups of researchers of any type, if the number of researchers is sufficiently large. If they are used at this aggregation level, the indicators are statistically robust and have a clear assessment meaning. Furthermore, they can be easily calculated and do not require any type of normalization.

Taking these characteristics into account, the peer assessment of countries and institutions is less convenient than the use of the double rank indicators. The weaknesses and strengths of peer assessment have been described (Wilsdon et al. 2015 p. 60), assuming that the peers are experts in the fields of the outputs under assessment. This requirement, which is mandatory, implies that the number of peers in a national research assessment has to be very high because research is currently highly specialized. Therefore, in addition to the economic costs of such an assessment, the time demanded from experienced researchers is unacceptable both as participants for preparing the documents requested for their own evaluations and as reviewers performing the assessments. The time required for these processes when added to that required for grant applications (Lawrence 2016) is too much for the most active scientists and weakens the research system.

Research assessments based on the double rank indicators do not require a single minute of researchers' attention, neither for an application, which is unnecessary, nor for an evaluation, which can be performed by an independent agency or governmental office. In addition, double rank indicators are fairly insensitive to the problems predicted by the Goodhart's law, "when a measure becomes a target, it ceases to be a good measure" (Strathern 1997 p. 308), because the only way to improve the indicators is to publish papers—not one or a few but all the papers—that attract more attention and receive a higher number of citations. Even in the improbable case that an institution or country was able to devise a procedure to add a certain number of spurious citations of each paper, the procedure would have a significant effect only for those papers with a low number of citations—it is unthinkable that a trick could be devised that significantly increased the number of citations of highly cited papers. Thus, the trick would affect the indicators very little because the fitting method of the power law is quite insensitive to increasing the number of citations of papers where that number is low (Section 5). In addition, the double rank power law would detect the described over-citation trick because it would introduce a deviation in the power law (Eq. 3). In fact, we detected different citation patterns to Chinese and Japanese papers in comparison with EU's and USA's papers (Section 5; Table 2), although these differences are small and are probably undetectable by other assessment methods.

A completely different issue is that, in contrast with expert assessments, bibliometric approaches do not measure the societal impact of research (Bornmann 2012, 2017). The solution to this problem is difficult, but the problem might be addressed by selecting the topics of the bibliometric searches. For example, a search in the research area of "plant sciences" will retrieve many papers that currently have little societal relevance but by adding the topics "drought" or "salinity," most of the retrieved papers will be about matters of societal relevance.

**10. Research evaluation in countries and institutions**

In most cases, the economic and research levels of countries are linked. Although innovation is a complex process that does not only involve progress in knowledge (Kline and Rosenberg 1986) and incremental technological innovations do not depend on research breakthroughs (Ettlie et al. 1984), research is a key element in the process of making markets (Miller and O'Leary 2007). Therefore, every country and institution should have an interest in improving its capacity to achieve technological breakthroughs, which implies that, as already highlighted, this capacity must be measured. However, for this purpose, research assessments must be performed in the research topics or fields that support technology and not in other fields of research (Brito and Rodríguez-Navarro 2018b; Rodríguez-Navarro and Brito 2018b).

Two indicators derived from the percentile-based double rank power law, the $e_p$ index and the expected breakthrough frequency, reveal two important characteristics of any research system. The former reveals an internal property of the system in comparison with the global one, which we have defined as *intrinsic efficiency* or *breakthrough potential* (Rodríguez-Navarro and Brito 2018b); the latter reveals the expected contribution to the global advancement of science or technology and depends on the $e_p$ index and also on the number of publications (Eq. (9)).

Although the mathematical properties on which the aforementioned indicators are based are the same for countries and institutions, the interpretation of their values in terms of research policy is different in the two cases.

*10.1. Evaluation of countries*

The most important research indicator for a country is probably the number of expected breakthroughs with reference to its size, which can be expected to vary, considering the high variation across countries of research efforts in terms of number of researchers and investments per million inhabitants (UNESCO 2016 p. 13). Therefore, the $P_{top\ 0.01\%}$ per million inhabitants of per million ECU of research investment seems to be a key indicator of the research capacity of the country. In contrast the $e_p$ index might be less informative, provided that it is reasonably high (e.g., > 0.11). In technologically competitive countries with many universities, a certain number of them might be heavily focused on teaching, which implies that their research is not very competitive. The $e_p$ index of these universities is low and although universities heavily focused on research have high $e_p$ index values, the $e_p$ index of the whole country will not be very high. This applies to the USA in technologies in which is a global leader (Brito and Rodríguez-Navarro 2018b; Rodríguez-Navarro and Brito 2018b); the same can be deduced from the data in Table 2 (in this case, it is worth recalling that $e_p = 10^{-\alpha}$).

*10.2. Evaluation of institutions*

The evaluation of institutions, in most cases universities, should be performed based only on the $e_p$ index. Normalization of the expected frequency of highly cited papers by size is complicated, because the size of a university is a fuzzy concept. Normalization in percentile-based indicators is in many cases performed by dividing by the total number of published papers (e.g., the Leiden Ranking), which is equivalent to calculating the probability of publishing a paper in a certain percentile (Eq. [3] and [4]). In the case of the 0.01 percentile ($P_{top\ 0.01\%}$) this probability is equal to $e_p$ raised to the fourth power (deduced from Eq. [9]).

In contrast with the mathematical simplicity of these evaluations, the interpretation of the results in terms of research policy is more difficult, because it might be necessary to consider the size and the research policy of the country to which the university belongs. For universities with a low $e_p$ index (< 0.1), the interpretation is simple: if they are research universities, they are barely competitive. In contrast, for universities with high $e_p$ index (e.g., > 0.15) the interpretation is more difficult. It is worth noting that, for a country's research policy, having four universities with an $e_p$ index of 0.15 is the same as having three universities with an $e_p$ index of 0.2, because breakthrough cumulative frequencies are additive. Therefore, a ranking of universities based on the $e_p$ index is enlightening when the universities belong to the same country but is more difficult to interpret when the ranking includes universities of many countries.

## 11. Conclusions

The indicators of research activity derived from the percentile-based double rank analysis have a strong theoretical foundation and they are simple, statistically sound and robust, and fairly stable over years. They are also easy to interpret, because their results can be expressed as probabilities or expected frequencies of breakthroughs. The $e_p$ index is an indicator that is independent of the size of the research system and the percentile of the evaluation, which reveals the *breakthrough potential* of any research system.

Furthermore, these new indicators are based on a percentile apportionment of publications that has been extensively used and studied. Another advantage is that no explicit field normalization is necessary because the double rank approach has an intrinsic normalization that eliminates the numbers of citations. Although the method can be formally applied to a mix of research fields, the analysis is imprecise if performed in this way; the best option is to apply the method to specific research fields or topics.

The double rank analysis settles a long-lasting discussion about whether scientific advancements lean on the shoulders of giants (Bornmann et al. 2010). The answer is that science progresses on the shoulders of the majority of scientists because giants would not exist without them. The notion that "an institution does not acquire an excellent score in an impact measurement due to the majority of scientists it employs and their publications, but due to its few very successful scientists or their small number of highly cited papers" (Bornmann 2017 p.781) does not apply if excellence is defined in terms of *breakthrough potential* (Rodríguez-Navarro and Brito 2018b). As a matter of fact, the probability that a publication from an institution reaches a high number of citations can be calculated from the citation distribution of the bulk of lowly cited papers.

Finally, an analysis of the pros and cons regarding the use of the double rank indicators in comparison with peer-based research assessments reveals that the former has many advantages, with the strict condition that it is not applied at low levels of aggregation.

**Acknowledgment**

This work was supported by the Spanish Ministerio de Economía y Competitividad, grant numbers FIS2014-52486-R and FIS2017-83709-R.